\documentclass[useAMS,usenatbib,letterpaper]{mn2e}
\usepackage{psfig}
\usepackage{times}
\usepackage{aas_macros}

\title[Star-formation activity as a function of dark matter
environment]
{\mbox{Linking star-formation and environment in the A901/902 supercluster}}
\author[M.E. Gray et al.]{M.E. Gray$^1$\thanks{email:
{\em meg@roe.ac.uk}}, C. Wolf$^{2,3}$, K. Meisenheimer$^3$,
A. Taylor$^1$, S. Dye$^4$, A. Borch$^3$,  M. Kleinheinrich$^3$\\ 
1. Institute for Astronomy, Blackford Hill, Edinburgh EH9 3HJ\\ 2. Department of
Physics, Denys Wilkinson Bldg., University of Oxford, Keble Road,
Oxford OX1 3RH\\ 3. Max-Planck-Institut f\"{u}r Astronomie,
K\"{o}nigstuhl 17, D-69117, Heidelberg, Germany\\ 4. Astrophysics
Group, Blackett Lab, Imperial College, Prince Consort Road, London SW7
2BW }
\begin{document}

\date{}

\pagerange{\pageref{firstpage}--\pageref{lastpage}} \pubyear{}

\maketitle

\label{firstpage}

\begin{abstract}
We investigate correlations between the location of galaxies in dense
environments and their degree of star-formation activity.  Using
photometric redshifts and spectral classifications from the unique
17-band COMBO-17 survey we are able to precisely isolate galaxies from
the Abell 901/902 supercluster within a thin redshift slice around
$z=0.16$.  We compare the detailed photometric properties of the
supercluster galaxies with the underlying dark matter density field as
revealed by weak gravitational lensing.  We find strong evidence for
segregation by type, with the highest density regions populated almost
exclusively by galaxies classified according to their rest-frame $U-V$
colours as quiescent.  We also observe a threshold surface mass
density from lensing, \mbox{$\kappa\sim 0.05$} (corresponding to a
physical density \mbox{$\Sigma = 2.5\times10^{14}h M_{\sun}$
Mpc$^{-2}$}), above which star-formation activity is rapidly
suppressed.  This abrupt transformation affects primarily the faint
end of the star-forming galaxy population and occurs at a local
surface number density corresponding to roughly $400h^2$ Mpc$^{-2}$
to a limit of $M^*_V+6$.  When only galaxies brighter than $M^*+1$ are
considered the trends with environment remain, but are more gradual
and extend beyond \mbox{$2h^{-1}$ Mpc} radius.
\end{abstract}


\section{Introduction}
At a given epoch, the observed properties of galaxies in dense
clusters of galaxies are known to vary from those residing in the more
rarefied field.  The classic example of such an environmental
dependence is the well-known morphology-density relation,
$\rm{T}-\Sigma$ \citep{dressler80,dressler97,treu03}, in which a
dramatic increase in the proportion of elliptical galaxies is seen
toward the cores of rich clusters.  Clusters of galaxies appear to act
to accelerate galaxy evolution, possibly through increased
galaxy-galaxy interactions and/or interactions between galaxies and
the global cluster environment.  It is possible to probe the
environmental dependence of galaxy evolution by tracing changes in
both structural parameters and star-formation activity.

Previous studies have found evidence for gradients in star-formation
properties \citep{abraham96,balogh97,hashimoto98} in samples of rich
clusters.  More recently a relation between star-formation activity
and local galaxy density (referred to here as the $\rm{SF}-\Sigma$
relation, analogous to the $\rm{T}-\Sigma$ relation) has been observed
in several guises: a break in $V-I$ colour in the outskirts of a
$z=0.4$ cluster \citep[][erratum in \citealt{kodama01err}]{kodama01}
and a decline in star-formation rate toward clusters in the 2dFGRS
\citep{lewis02} and with increasing galaxy density in the SDSS
\citep{gomez03}. Intriguingly, all three of these studies find
evidence for threshold local galaxy densities above which
star-formation activity is rapidly quenched.  The range of densities
probed suggests that environmentally driven changes in the
star-formation rate occur not only in the cores of rich clusters but
also in regions more associated with poorer groups and infall regions.
This picture is consistent with hierarchical models of structure
formation \citep[e.g.][]{kauffmann93} in which clusters grow by
continuous accretion of galaxies or groups of galaxies from the field.
Less clear, however, is the mechanism responsible for driving these
transformations \citep{balogh00,diaferio01,okamoto03} and whether the
same mechanism orchestrates changes in observed morphological,
spectral, and photometric properties on similar timescales.

\begin{figure}
\psfig{figure=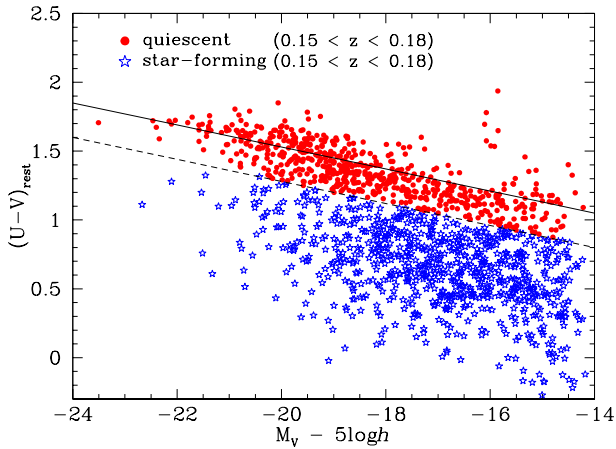,width=0.98\columnwidth}
\caption{Rest-frame $U-V$ vs. absolute $V-$band colour-magnitude
diagram for supercluster galaxies with $0.15<z_{\rm phot}<0.18$. The
solid line indicates a fit to the cluster red-sequence with fixed
slope of -0.08, and the dashed line indicates the adopted
$\Delta\left(U-V\right)_c$ cutoff between quiescent and star-forming
galaxies.}
\label{fig-selectionplot}
\end{figure}

The combination of wide-field gravitational lensing and multi-band
imaging offers a new approach to the study of galaxy evolution.  Weak
lensing directly probes the underlying density field (which may or may
not be well correlated to local galaxy density), and thus offers a
different but complementary measure of environment.  The key quantity
to consider is the total projected surface mass density, $\kappa$,
expressed in units of the critical density for lensing.  The
construction of a $\rm{SF}-\kappa$ relation and its comparison with
the analogous $\rm{SF}-\Sigma$ relation may shed new light on the
mechanisms responsible for the transformation of galaxies as they are
accreted onto a dense environment.  In this paper we exploit the
unique dataset provided by the COMBO-17 survey and combine photometric
redshifts, spectral energy distributions (SEDs) and weak gravitational
lensing to investigate correlations between star-formation activity
and environment in a supercluster at $z=0.16$.  Throughout, we assume
$\Omega_m=0.3, \Omega_\Lambda=0.7,$ and \mbox{$H_0=100h$ km s$^{-1}$
Mpc$^{-1}$}.

\begin{figure*}
\psfig{figure=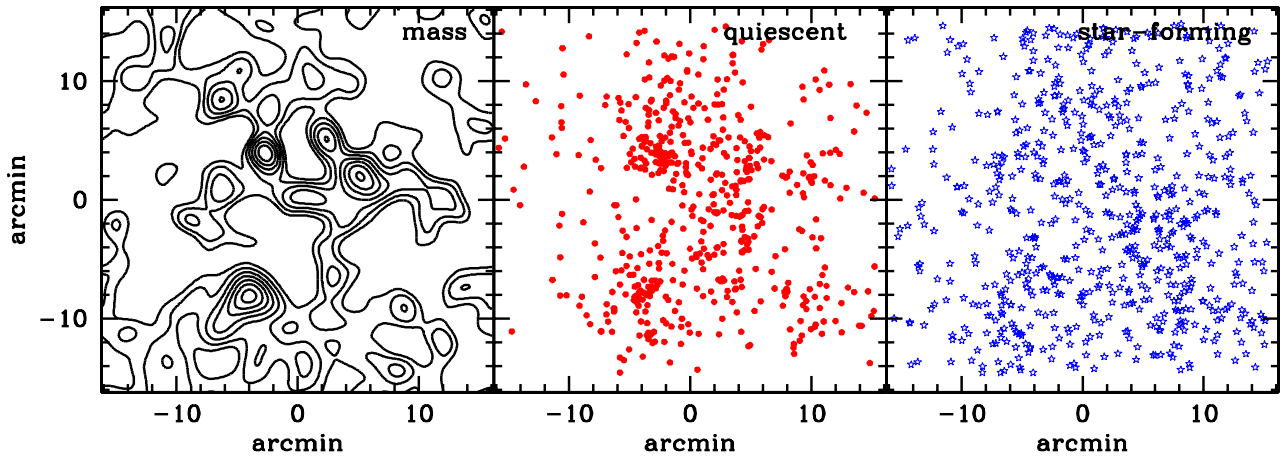,width=0.98\textwidth}
\caption{Distribution of redshift-selected supercluster galaxies
relative to lensing mass map of \citealt{gray02} (left).  The
contours represent a spacing of $\Delta\kappa=0.02$, smoothed with a
Gaussian of $\sigma=60\arcsec$.  Note the decreasing concentration of
the galaxies with increasing degree of star-formation activity.}
\label{fig-distributions}
\end{figure*}

\section{Observations and galaxy classifications}
The A901/902 supercluster was observed in 17 bands (5 broad-band, 12
medium-band) with the $0.5^{\circ} \times 0.5^{\circ}$ ESO/MPG
Wide-Field Imager as part of the COMBO-17 spectrophotometric survey
\citep{wolf03}.  The supercluster consists of three main cluster-sized
mass concentrations all at $z=0.16$ and contained within the field of
view of the instrument.  Preliminary results from a subset of the full
dataset, consisting of the broad-band $B-$ and $R-$band data alone,
were presented in \citet{gray02}.  These findings included a 2-D map
of the dimensionless projected surface mass density, $\kappa$, which
was the result of a weak gravitational lensing reconstruction based on
the deep $R$-band image.  The complex structure and obvious
non-relaxation of the system make it an ideal laboratory for probing a
range of environments.

The procedure for classification and redshift estimation has been
described in detail several times elsewhere
\citep{wolf03,bell03,wolf01b}.  Here, we only stress that on average
we expect galaxy redshifts to be accurate to \mbox{$\sigma_z \approx
0.03$} for a random sample (with early-types good to around $\sim
0.01$), while faint starburst galaxies can be worse than $\sim 0.05$.
The 17-band system of `fuzzy spectroscopy' allows for detailed
rendering of template fits.  With knowledge of the SED shape we can
directly reconstruct rest-frame luminosities in Johnson $U$- and
$V$-bands across a wide redshift range without generic $k$-correction.

Armed with both SED classifications and photometric redshifts, it is
possible to accurately isolate those galaxies belonging to the
foreground lensing structure.  Having first selected galaxies with
$R<24$ and $0.15<z<0.18$, we then further subdivide the supercluster
population into two classes according to photometric properties.
Fig.~\ref{fig-selectionplot} shows the measured rest-frame $U-V$
vs. absolute $V-$band colour-magnitude diagram for the
redshift-selected galaxies.  As in the study of the evolution of 5000
early-type galaxies in the COMBO-17 survey from $0.2<z<1.1$
\citep{bell03}, we observe a distinctly bimodal distribution.  We fix
the slope of the prominent colour-magnitude relation (CMR) as -0.08
\citep{bower92} and fit the intercept for all galaxies with
\mbox{$M_V<-18 +5\log h$}.  We next define an offset \mbox{$\Delta
(U-V)_c=0.25$} to mark the division between our two classes of
galaxies.  Redward of this limit we have the prominent ridge of
red-sequence early-type galaxies, and blueward is the second
population of star-forming galaxies.  Note that this division between
`quiescent' and `star-forming' galaxies is made purely on photometric
information with no considerations regarding morphology.

This method of selecting cluster members by photometric redshift is in
contrast to the alternative approach of selecting the red sequence
along the $B-R$ colour-magnitude relation \citep[cf.][]{gray02}.
While in that case blind red-sequence selection alone managed to
isolate many of the quiescent galaxies, it was insensitive to the
large number of blue star-forming galaxies within the structure, and
furthermore suffered from the inclusion of galaxies along the line of
sight.  The combination of redshifts and photometric classifications
offers a much more precise characterization of the luminous properties
of the cluster, while simultaneously eliminating confusion from
foreground/background contamination.  Furthermore, we note that
increasing the range for redshift selection to $0.15<z<0.2$ makes
negligible difference to the following analysis and conclusions.  This
consistency implies that the photometric redshifts of the quiescent
galaxies are indeed accurate to $\sigma_z\sim0.01$

\begin{figure}
\psfig{figure=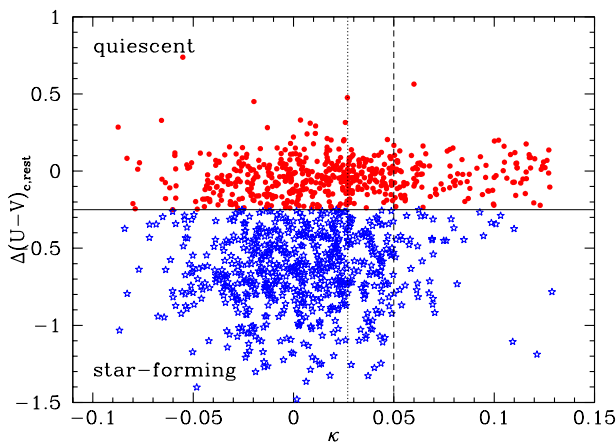,width=0.98\columnwidth} 
\caption{Star-formation activity {\it vs.} local surface mass density
$\kappa$ for each of the supercluster galaxies in the range
$0.15<z_{\rm phot}<0.18$.  A strong segregation by type is seen, with
the highest density regions populated almost exclusively by those
galaxies classified as quiescent.  The dotted line indicates the
$1\sigma$ noise level in the mass map.  The dashed line at
$\kappa\sim0.05$ illustrates the mass threshold above which the number
of star-forming galaxies is rapidly reduced.}
\label{fig-sedkappa}
\end{figure}

\begin{figure}
\psfig{figure=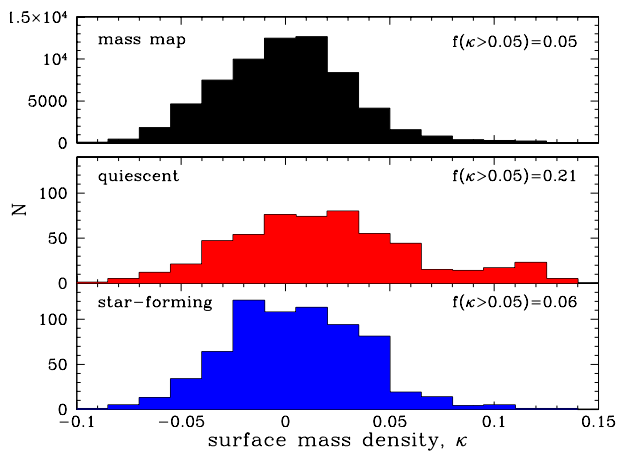,width=0.98\columnwidth,angle=0}
\caption{Distribution of environments populated by quiescent and
star-forming galaxies, with surface mass density sampled at the
location of each supercluster galaxy.  The fraction of each population
residing in high-density regions is shown. The top histogram
illustrates the distribution of densities from the lensing mass-map.}
\label{fig-kappabins}
\end{figure}

\section{Galaxy properties and environment}

Fig.~\ref{fig-distributions} shows the distribution of the
redshift-selected galaxies relative to the mass map of \citet{gray02}.
A segregation effect by type is clearly evident, with the quiescent
galaxies (Fig.~\ref{fig-distributions}b) much more strongly
concentrated than the actively star-forming galaxies
(Fig.~\ref{fig-distributions}c).  The mass map
(Fig.~\ref{fig-distributions}a) was constructed using a sample of
quiescent galaxies chosen to have $B-R$ colours redder than the
colour-magnitude sequence, so overlap between the lensing and lensed
populations here will be negligible.  The map was smoothed with a
Gaussian of $\sigma=60$ arcsec ($\sim120 h^{-1}$ kpc).  The rms noise
in the map was estimated to be $\sigma_\kappa=0.027$ by constructing 32
realizations of noise maps made by randomizing the shapes of the
galaxies in the background catalogue.  Thus, the three cluster peaks
are each detected at the $>3\sigma$ level, but mass peaks
corresponding to additional concentrations of galaxies (i.e. a group in
the SW corner of the field) are not significantly detected above the
noise.

We find further evidence for the segregation of galaxy populations and
a type-dependent correlation with the mass distribution by sampling
the lensing mass map at the location of each of the supercluster
galaxies.  Using the deviation blueward of the cluster CMR sequence as
a measure of star-formation activity, Fig.~\ref{fig-sedkappa} reveals
that the highest density regions are populated almost exclusively by
those galaxies classified as `quiescent'.  Furthermore, there appears
to exist a critical surface mass density, $\kappa\sim 0.05$, above
which few star-forming galaxies are seen (although clearly a small
number of galaxies would be expected to appear in projection against
the mass peaks).  We acknowledge that the noise in the mass
reconstruction could scatter galaxies from lower to apparently higher
densities, so this value may in reality be an upper limit.  However,
as will be seen later in the paper, obervations of a similar threshold
in the galaxy number density (a less noisy quantity) and a correlation
between number density and mass density supports the idea that the
mass threshold is not drastically lower.

The strikingly different density distributions for each population are
shown in Fig.~\ref{fig-kappabins}.  The distribution of pixels in the
mass map itself is also shown: the noise inherent in such mass
reconstructions is reflected in the number of `negative mass' pixels,
but we note the asymmetry of the distribution and the tail at high
densities.  The fraction of star-forming galaxies found in regions
with density above this threshold, $f(\kappa>0.05)$, is 0.06, compared
to 0.21 for quiescent galaxies.  The abrupt transition to a population
dominated by passive galaxies at high densities indicates that an
environmental effect may be responsible.  It is possible that we are
observing the truncation of active star formation as galaxies are
accreted onto the dense supercluster environment.

\begin{figure*}
\centerline{\psfig{figure=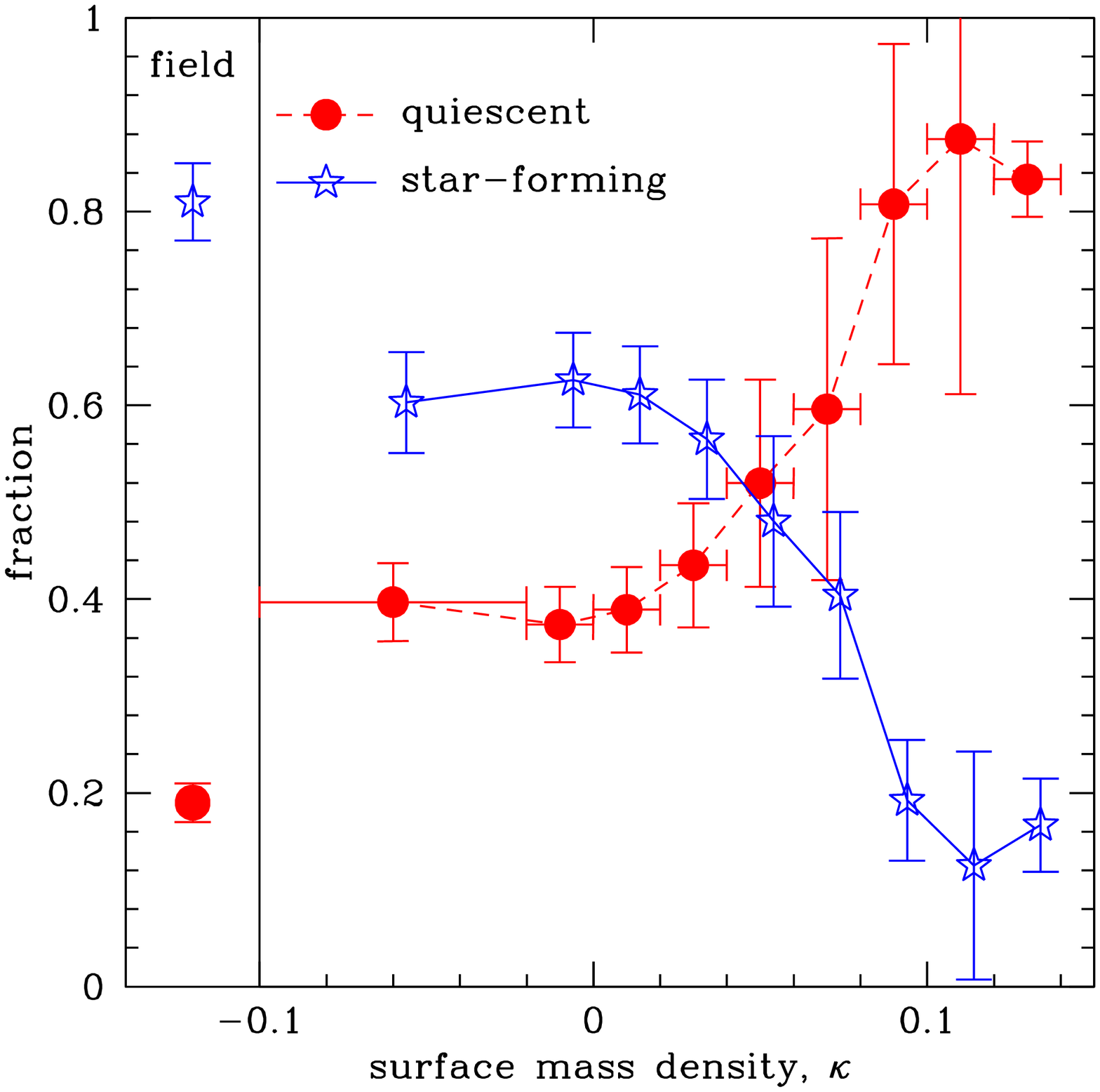,width=0.34\textwidth,angle=0}
\psfig{figure=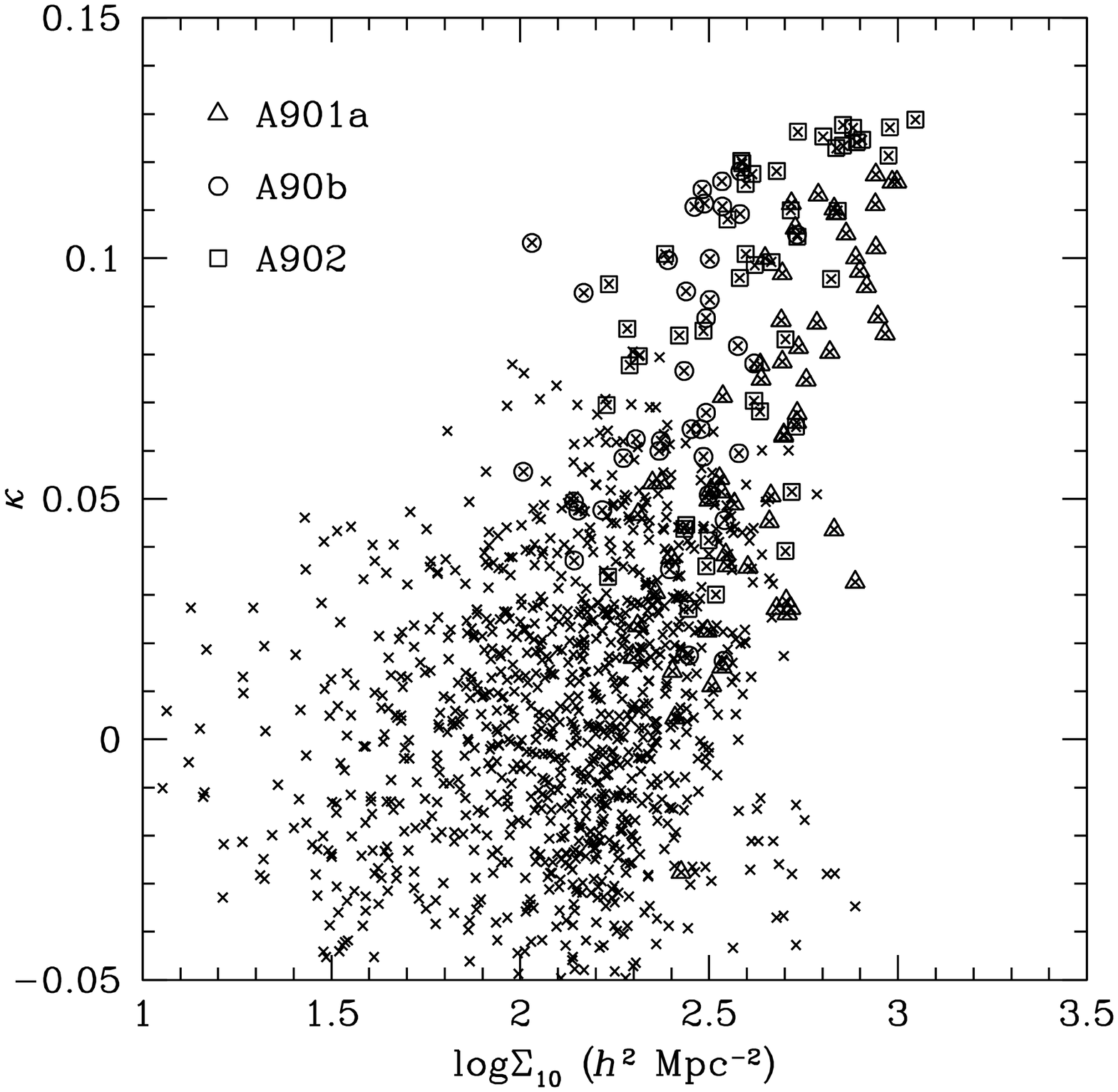,width=0.34\textwidth,angle=0}
\psfig{figure=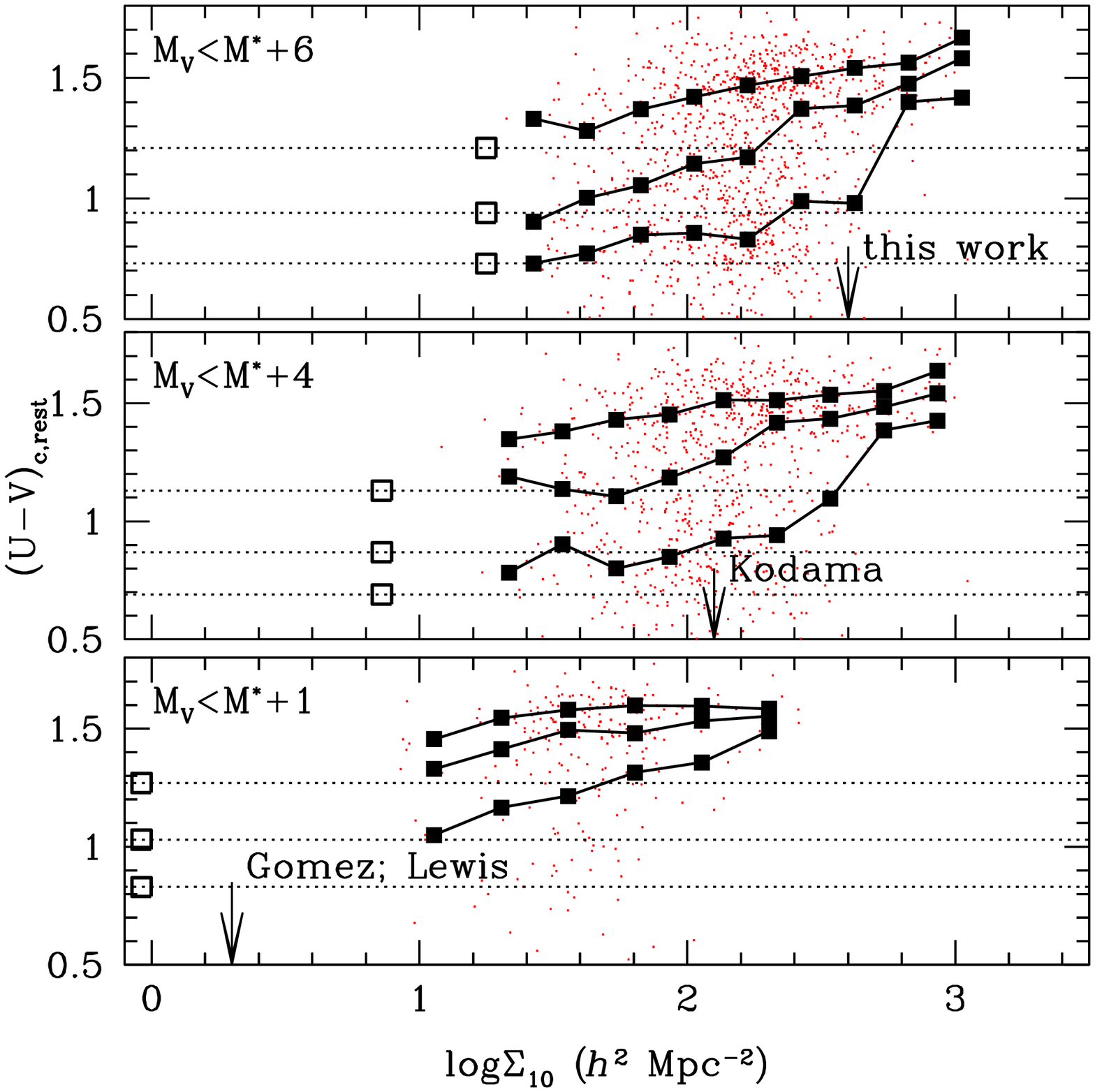,width=0.34\textwidth,angle=0}}
\caption{{\em Left:} SF-$\kappa$ relation.  A dramatic change in
relative abundance of each galaxy type at a given density is seen as
the proportion of star-forming galaxies declines and quiescent
galaxies become more dominant in dense regions.  The symbols at left
show the corresponding fractions reached in the field, obtained from
an identical redshift slice within two blank COMBO-17 fields.  As
negative values for $\kappa$ indicate noise in the mass
reconstruction, these data have been collected into a single bin.
Horizontal error bars represent bin widths and are shown for the
quiescent galaxies only, for clarity.  {\em Centre:} Local mass
density vs. local galaxy number density, sampled at the location of
supercluster galaxies.  A global trend is evident but with
considerable scatter from cluster to cluster (also note the
logarithmic horizontal axis). {\em Right:} Effect of magnitude limit
on galaxy colour trends.  The rest-frame $(U-V)_c$ colour is shown as
a function of local galaxy density for various limiting magnitudes to
mimic relevant studies in the literature.  The corresponding `critical
densities' for the suppression of star-formation are indicated by
arrows. The solid lines show the $25^{th}$, $50^{th}$, and $75^{th}$
percentiles of each distribution, and in each case the trends are
strongest for the bluest quartile.  The dotted lines represent the
same percentiles observed in the COMBO-17 offset fields for each
limiting magnitude, and the open squares show the value of
$\Sigma_{10}$ at which those quantities are measured.}
\label{fig-seddensity}
\end{figure*}

When considering the relative abundances of each type of galaxy at a
given density, we find that behaviour analogous to the traditional
morphology-density relationship \citep{dressler80} is evident in the
${\rm SF}-\kappa$ diagram in Fig.~\ref{fig-seddensity}a.  The fraction
of galaxies that are actively forming stars plummets at high surface
mass densities, with a compensating rise in the fraction of quiescent
galaxies.  For comparison, the relative abundances of each type in the
field were determined from identical redshift slices in two blank
COMBO-17 fields.  Even at low surface mass densities the fraction of
galaxies classified as quiescent in the supercluster is in excess of
the field value.  However, measurements in this regime may suffer from
contamination by low-mass groups whose mass peaks are not detected
above the noise, but which nonetheless contain a higher-than-field
fraction of quiescent galaxies.  

Considering the threshold density of $\kappa=0.05$, we convert to
physical units by assuming that the source galaxies for lensing reside
on a plane at $z=1.0$ \citep[cf.][]{gray02,brown03}.  Then the
critical density for a lens at $z=0.16$ is $\Sigma_{\rm crit} =
5.0\times10^{15}h M_{\sun}$ Mpc$^{-2}$, which yields a projected
density of $\Sigma_\kappa=\kappa \Sigma_{\rm crit} = 2.5\times10^{14}h
M_{\sun}$ Mpc$^{-2}$.  The advantage of gravitational lensing is that
this measure of total projected mass has the potential to allow direct
comparison with dark matter particles in simulations, without
assumptions regarding if and how light traces mass.  In addition, most
of the mass in the inner regions of clusters is thought to be smoothly
distributed, with only $10-20\%$ of the total mass in the core regions
associated with individual galaxy haloes \citep{natarajan02}.
However, in order to compare these results with previous studies of
galaxy evolution in dense environments, we then ask: does this total
mass density correspond to a counterpart threshold galaxy number
density?

\cite{gray02} found that the distribution of (colour-selected)
early-type galaxies alone relative to the mass distribution did not
scale in the same way for each of the three clusters.  Indeed, in one
case the two distributions showed significant misalignment.  However,
with the addition of the full redshift-selected sample of supercluster
galaxies available for this work we find a qualitative improvement
(though still with large scatter) in the relation between surface mass
density and local galaxy number density.  We define $\Sigma_{10}$ as
the local density by calculating the area of the circle containing a
given galaxy and its ten nearest neighbours within our redshift slice.
We exclude all galaxies for which such a circle would extend outside
the field boundaries.  Fig.~\ref{fig-seddensity}b shows that the
surface mass density $\kappa\sim0.05$ corresponds to a local galaxy
density of $1.8<\log\Sigma_{10}<2.8$, though this sampling is
restricted to the location of the supercluster galaxies and so cannot
probe areas containing dark matter but few galaxies.  Therefore, though
local number density is not necessarily a direct proxy for mass
density, Fig.~\ref{fig-seddensity}b shows that it serves here as a
first approximation.

To compare this galaxy density threshold with previous studies, we
note that our limit of $R<24$ for photometric redshift determination
corresponds to an absolute $V-$band magnitude of \mbox{$M_V=-14.3 +
5\log{h}$}.  Taking \mbox{$M^*_V=-20.2+5\log{h}$} \citep{brown01},
this corresponds to $M^*_V+6$, or $0.004L^*$.  While the limiting
apparent magnitude is similar to that of \cite{kodama01}, the lower
redshift of the A901 supercluster relative to their cluster at $z=0.4$
allows us to probe two magnitudes deeper down the luminosity function.
Thus by applying the relevant luminosity limits we can make direct
comparisons with the threshold densities quoted by \citealt{kodama01}
(corrected from $\sim$400$h^2$ to $\sim$120$h^2$ Mpc$^{-2}$, brighter
than $M^*_V+4$ in \citealt{kodama01err}) and \citealt{lewis02} and
\citealt{gomez03} ($\sim$2$h^2$ Mpc$^{-2}$, brighter than $M^*+1$).
\citeauthor{gomez03} noted that even accounting for the change in
luminosity limit these two measures appear to differ by an order of
magnitude.  Even following correction of the \citeauthor{kodama01}
densities in the subsequent erratum, the breaks in galaxy properties
still occur at different densities and different clustercentric radii.
With the COMBO-17 dataset, we are in a position to resolve this
anomaly.

Fig.~\ref{fig-seddensity}c shows the rest-frame colour trends with
local galaxy density for all three magnitude limits under
consideration ($M^*_V+6, M^*_V+4, M^*_V+1$).  In each case the bluest
quartile of the galaxy population shows the strongest change with
density, in agreement with the previous studies.  For the full dataset
(top panel), a sharp transition to redder colours is visible at
\mbox{$\log\Sigma_{10}\sim2.6$} or $\sim$400$h^2$ Mpc$^{-2}$, as
anticipated from the $\kappa-\log{\Sigma_{10}}$ relation.  If we apply
the \citeauthor{kodama01} $M^*+4$ luminosity limit (middle panel), the
break is less abrupt but a change in colour does appear to occur close
to their critical density.  However, when we restrict ourselves to the
brighter limit (and thus lower galaxy densities) of \cite{gomez03} and
\cite{lewis02}, the trend with density is smoother.

Nevertheless, for all three magnitude limits the rest-frame colour is
still redder than field values at all densities.  These field
densities and colour quartiles, taken from the same $0.15<z<0.18$
redshift slice in the COMBO-17 offset fields, are indicated by the
open squares and dashed lines in Fig.~\ref{fig-seddensity}c.  An
extrapolation of each curve back to the corresponding density in the
field for that limiting magnitude shows good agreement with the
measured colours at low densities.  Furthermore, we note that the
density of field galaxies in COMBO-17 at the bright limit ($M^*_V+1$)
is close to the critical densities of \cite{gomez03} and
\cite{lewis02}, but that the colour transition for bright galaxies is
noticably smoother. Hence, this observed density threshold of $\sim 2
h^2 {\rm Mpc}^{-2}$ could more appropriately be interpreted as the
point at which the abundance ratios begin to depart from those
observed in the field.  While at fainter limits this departure begins
at densities $\Sigma_{10}\sim10h^2 {\rm Mpc}^{-2}$, this is in
addition to the secondary and sharper cut-off seen here and in
\cite{kodama01}.  This additional threshold occurs at a density which
is an order-of-magnitude above the field density, and affects
primarily the faint end of the luminosity function.

\section{Conclusions}
In summary, by adjusting the luminosity limits of our sample we
observe two distinct effects: a gradual change in the properties of
bright galaxies from several virial radii, and an abrupt change in the
properties of the faint (starburst) population, occurring at
\mbox{$r<0.5h^{-1}$ Mpc} from the centre of each cluster.  For the
first time we are able to measure environmental effects using not only
galaxy density but also direct measurements of the projected
surface mass density $\kappa$ from weak gravitational lensing.  The
observed galaxy density thresholds and clustercentric distances appear
to be in broad agreement with both the low- and high-redshift studies
cited above, when the differing luminosity limits are taken into
account.  We postulate a simple scenario in which we may be observing
the effects of multiple mechanisms influencing the transformation of
the bright and faint galaxies in the low- and high-density regimes.

In the transitional or infall regions where densities are relatively
low, gradual processes over timescales of several Gyr are dominant.
The most likely mechanisms for the slow suppression of star formation
include galaxy `harassment' in the form of high-speed galaxy-galaxy
interactions \citep{moore96}, or `suffocation' following the removal
of a diffuse hot gaseous reservoir and the subsequent consumption of
any remaining gas \citep{larson80}.  At higher densities (or lower
clustercentric radii) the abrupt observed transition may indicate the
threshold at which the influence of the cluster potential takes hold.
This influence may manifest itself through tidal truncation of galaxy
haloes \citep{merritt83, merritt84,natarajan02} or by ram-pressure
stripping of galactic gas \citep{gunn72}.  Ram-pressure stripping
requires dense gas so for a typical massive cluster becomes
ineffective beyond $\sim$1 Mpc \citep{treu03}; likewise, the tidal
effects of the cluster are felt over a very narrow range (100-200 kpc)
and are most effective at disrupting less dense galaxies
\citep{treu03,moore98}.

For this particular supercluster field, the COMBO-17 dataset has been
further augmented by 2dF spectra and recent deep X-ray observations
with XMM-Newton. Both additions will further test the mechanisms
responsible for the galaxy transformations within this system.
Analysis of the detailed spectroscopy of 300 supercluster galaxies
will reveal their star formation histories, in particular signatures
of cluster-induced starbursts or recently truncated star formation.
Using measurement of emission and absorption features we will trace
the distribution of starburst and post-starburst galaxies relative to
both dark matter environment and local galaxy density.  The X-ray data
will probe the hot ICM and the test the effectiveness of ram-pressure
stripping in these regimes.  In addition, the remaining COMBO-17
fields cover a wider mass spectrum, including multiple cluster
systems, an isolated cluster, and blank fields.  Applying the joint
lensing and multi-colour analysis to all these fields will help to
untangle environmental links between dark matter distribution, local
galaxy number density, and galaxy evolution.

\section*{Acknowledgments}

MEG acknowledges a PPARC Postdoctoral Fellowship and thanks Ian Smail
and Eelco van Kampen for useful discussions.  CW was supported by the
PPARC rolling grant in Observational Cosmology at the University of
Oxford.  SD was supported by a PPARC rolling grant at Imperial
College.  We are grateful to the anonymous referee for useful
suggestions that improved the paper.


\bsp

\label{lastpage}

\end{document}